\documentclass[a4paper,11pt]{article}
\pdfoutput=1 

\usepackage{jinstpub} 

\usepackage{siunitx}
\usepackage{mhchem}
\usepackage{natbib}
\usepackage[shortlabels]{enumitem}
\usepackage{hyperref}

\graphicspath{ {./images/} }

\title{Production and characterization of random electrode sectorization in GEM foils}

\author[a,1]{A. Pellecchia,\note{Corresponding author.}}
\author[b]{M. Bianco,}
\author[b]{R. De Oliveira,}
\author[c]{F. Fallavollita,}
\author[d]{D. Fiorina,}
\author[d]{N. Rosi}
\author[a]{and P. Verwilligen}
\author{on behalf of the CMS Muon group}

\affiliation[a]{INFN sezione di Bari, Italy}
\affiliation[b]{CERN, Switzerland}
\affiliation[c]{Max Planck Institut für Physik - Werner Heisenberg Institute, Germany}
\affiliation[d]{Università degli studi di Pavia, Italy}

\emailAdd{antonello.pellecchia@cern.ch}

\abstract{
  In triple-GEM detectors, the segmentation of GEM foils in electrically independent sectors allows reducing the probability of discharge damage to the detector and improving the detector rate capability.
  However, a segmented foil presents thin dead regions in the separation between two sectors and the segmentation pattern has to be manually aligned with the GEM hole pattern during the foil manufacturing, a procedure potentially sensitive to errors.

  We describe the production and characterization of triple-GEM detectors obtained with an innovative GEM foil segmentation technique, the ``random hole segmentation'', that allows easier manufacturing of segmented GEM foils.
  The electrical stability to high voltage and the gain uniformity of a random-hole segmented triple-GEM prototype are measured.
  The results of a test beam on a prototype assembled for the Phase-2 GEM upgrade of the CMS experiment are also presented.
  A high statistics efficiency measurement shows that the random hole segmentation can limit the efficiency loss of the detector in the areas between two sectors, making it a viable alternative to blank segmentation for the GEM foil manufacturing of large-area detector systems. 
}


\proceeding{The 7$^{\text{th}}$ International Conference on Micro Pattern Gaseous Detectors\\
Weizmann Institute of Science, Rehovot, Israel\\
December 11-16, 2022}

\notoc
\begin{document}
\maketitle
\flushbottom

\section{Introduction}
\label{sec:intro}

Triple-GEM detectors are a consolidated technology among micro-pattern gaseous detectors (MPGDs), chosen for several high-rate applications for tracking at high energy physics experiments due to their high rate capability and good space resolution \cite{sauli_gem}.

In several applications of GEM detectors in experiments, the top and bottom electrodes of the GEM foils are segmented in electrically independent sectors.
The segmentation is essential for high-rate applications to limit the capacitance of electrodes facing each other, thus decreasing the energy released during a discharge. 
Additionally, lowering the area of each sector reduces the current induced during irradiation on the electrodes by the avalanche charges, which flows through the high voltage protection resistors and limits the detector rate capability.

The standard ``blank'' segmentation is added in a GEM foil as a region free of hole pattern in the etching mask.
After the GEM foil etching, the copper electrodes in the region between two sectors are further etched, leaving an exposed polyimide region.
The two main downsides of segmented foils are then:
\begin{itemize}
  \item the introduction of thin dead regions in the separation between two sectors;
  \item the segmentation pattern has to be manually aligned with the GEM hole pattern during the GEM foil manufacturing, a procedure potentially sensitive to errors.
\end{itemize}

The mitigation of these issues is expecially important for high-rate systems such as the ME0 system of the Compact Muon Solenoid (CMS) Phase-2 Upgrade at the LHC \cite{muon_tdr}:
the ME0 detector station, scheduled for installation in the CMS muon spectrometer in the LHC Long Shutdown 3 (LS3), will be made of 18 stacks of triple-GEM detectors for each of the two endcaps.
Each GEM foil in the ME0 detectors will be segmented along the azimuthal direction with respect to the LHC beam line in 40 sectors \cite{me0_ratecapability}.
The segmentation has been optimized to reduce the gain drop under irradiation by the LHC background due to rate capability effects, but it is expected to lower the geometrical acceptance of the detector by approximately 4\% due to the addition of a dead area.

An alternative manufacturing technique, the ``random hole segmentation'', consists of maintaining the hole pattern in the separation between sectors, thus simplifying the manufacturing process.
Previous studies \cite{florian} also hinted that, thanks to a higher foil transparency, detectors with random-hole segmented GEM foils might partially restore the efficiency loss in the segmentation region.

The following sections describe the production of a random-hole segmented GEM foil and the characterization of a triple-GEM prototype instrumented with such foils.
Finally, preliminary test beam results of a random-hole segmented detector for the ME0 station are presented and its efficiency for muons is compared with the one of a blank-segmented triple-GEM prototype.

\section{Production of random hole segmented GEM foil}

The detailed steps of the production of GEM foils with random hole segmentation using the single-mask technique are represented in Fig.\,\ref{fig:random_hole_production} as adapted from \cite{michele_random}:

\begin{figure}[tb]
    \centering
    \includegraphics[width=\textwidth]{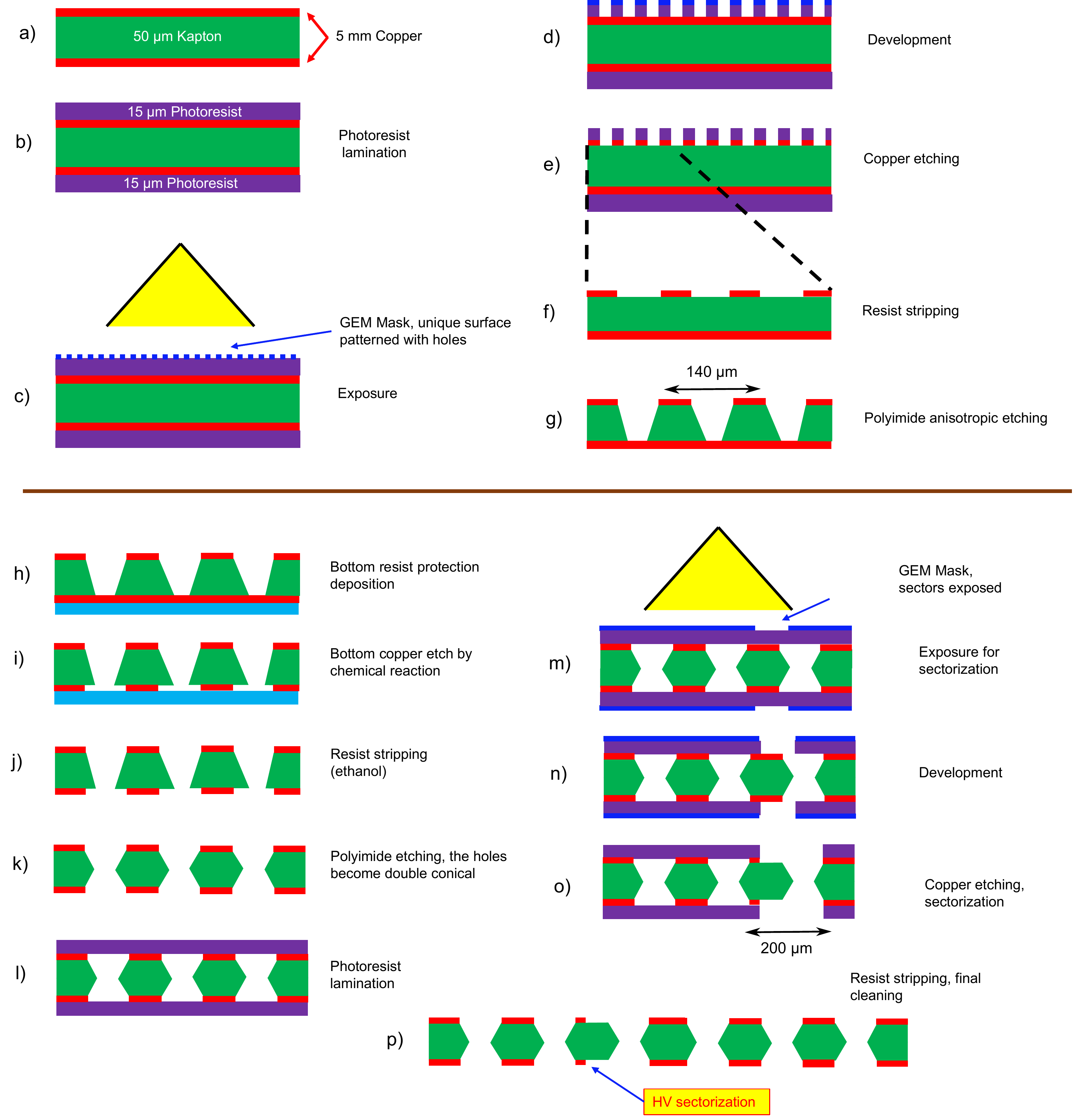}
        \caption{Production steps for a GEM foil with random hole segmentation \cite{michele_random}. Not to scale.}
    \label{fig:random_hole_production}
\end{figure}

\begin{enumerate}[a)]
        \item the production starts from a \SI{50}{\micro\m} thick polyimide (PI) foil covered by two \SI{5}{\micro\m} thick copper layers;
        \item two photoresist layers are applied on top and bottom;
        \item a mask with the hole pattern is applied on the top photoresist layer (in the blank segmentation foil production, this mask also presents the segmentation pattern);
        \item the mask is exposed to UV light and the pattern is developed on the top photoresist;
        \item the copper is etched in a chemical bath;
        \item the photoresist layer is removed;
        \item the PI is etched from the top, giving rise to a conical hole pattern;
        \item the bottom copper layer is laminated by a protective layer;
        \item the bottom copper layer is etched;
        \item the resist layer is removed;
        \item the polyimide is etched from the bottom layer, leaving double-conical holes;
        \item to apply the segmentation, the foil is covered again by a photoresist layer on both sides;
        \item a mask with the segmentation pattern is applied;
            \begin{itemize}
                \item since the original mask at step c) did not have a segmentation pattern, no alignment is required and the mask may also cut a GEM hole;
                \item this step is also followed in the blank segmentation production, but the segmentation mask would need to be aligned with the pattern already present in the foil;
            \end{itemize}
        \item the mask is developed, leaving the copper exposed;
        \item the copper in the segmentation zones is etched, leaving the polyimide exposed;
        \item the photoresist layer is removed.
\end{enumerate}

\section{\boldmath Gain uniformity measurement} 

  A triple-GEM prototype with random hole segmentation made of 20$\times$\SI{10}{\centi\m\squared} GEM foils was assembled and tested in laboratory.
  The GEM foils were segmented in four rectangular sectors along the larger side.

  The prototype was tested for electrical stability of the GEM foils \cite{ciccio_random_hole}, with no instabilities observed up to a working point corresponding to an effective gain of \SI{2e4}{} -- similar to the typical stability range of detectors with blank-segmented foils. 
  A fine-grain uniformity map of the effective gain with \SI{2}{\milli\m} pitch has been measured with a \ce{^{55}Fe} source moved by a stepper motor, while measuring the anode current on the strips with a picoammeter. 
  The resulting uniformity map (Fig.\,\ref{fig:20x10_efficiency} left) shows a gain loss between 20 and 30\% in the sector separation lines with respect to the center of the GEM sector.


\section{Efficiency uniformity measurement} 

\begin{figure}[tb]
      \centering
      \includegraphics[width=\textwidth]{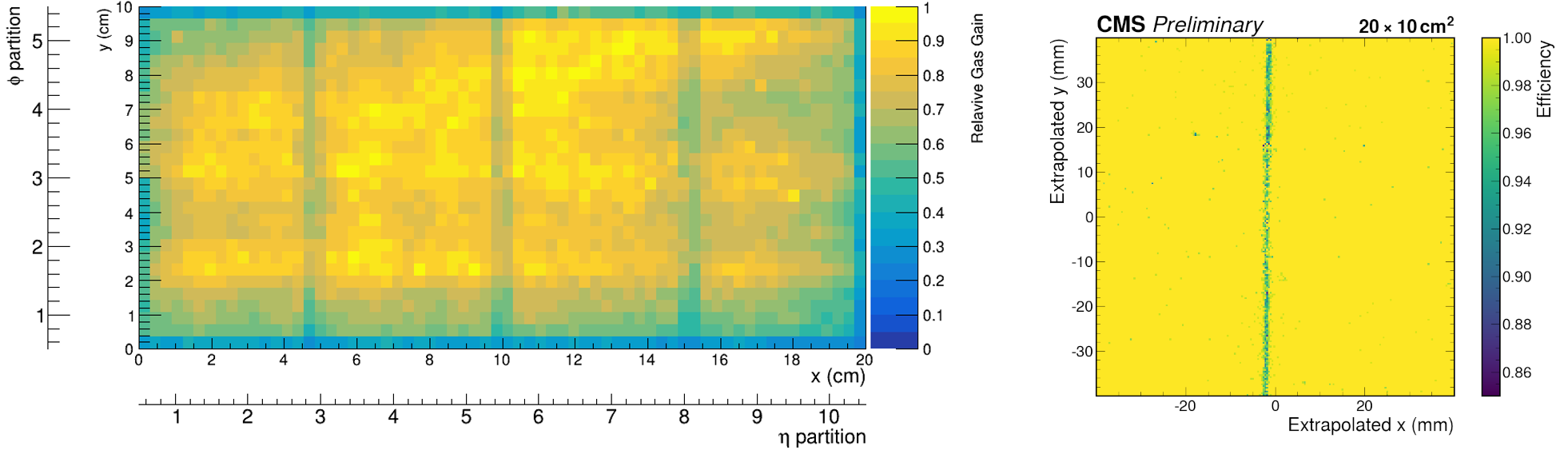}
      \caption{On the left, gain uniformity map of the 20$\times$10 prototype measured with a movable source \cite{ciccio_random_hole}.
      On the right, efficiency map of the 20$\times$10 random-hole segmented prototype.}
      \label{fig:20x10_efficiency}
  \end{figure}

  The $20\times\SI{10}{\centi\m\squared}$ prototype has been tested with $\SI{80}{\giga\eV}/\text{c}$ muons in test beam at the CERN North Area.
  Using a two-dimensional triple-GEM tracker with point space resolution \SI{75}{\micro\m}, a high-granularity efficiency map of the detector is obtained by binning the propagated track position by a few hundreds of \SI{}{\micro\m} (Fig.\,\ref{fig:20x10_efficiency} right).
  The map shows a very limited efficiency loss in the sectorization line, with an efficiency dip of only 4\%.

  To compare the impact of the random hole segmentation on the efficiency, a CMS ME0 detector instrumented with random-hole segmented GEM foils was operated in test beam together with a blank-segmented ME0 detector.
  ME0 detectors have trapezoidal shape (with approximate bases of 23 and \SI{46}{\centi\m} and height \SI{73}{\centi\m}) and are segmented in 40 sectors along the vertical direction.

\begin{figure}[tb]
      \centering
      \includegraphics[width=.495\textwidth]{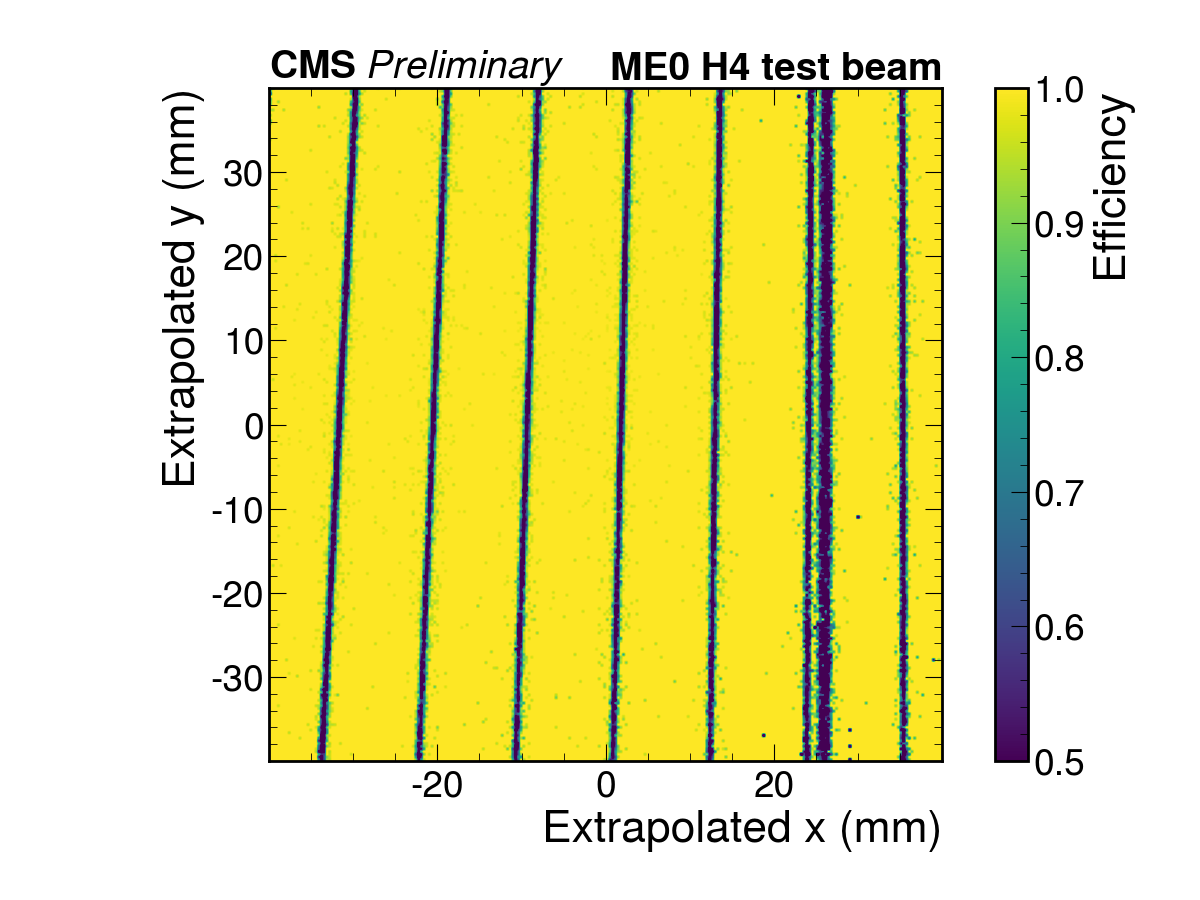}
      \includegraphics[width=.495\textwidth]{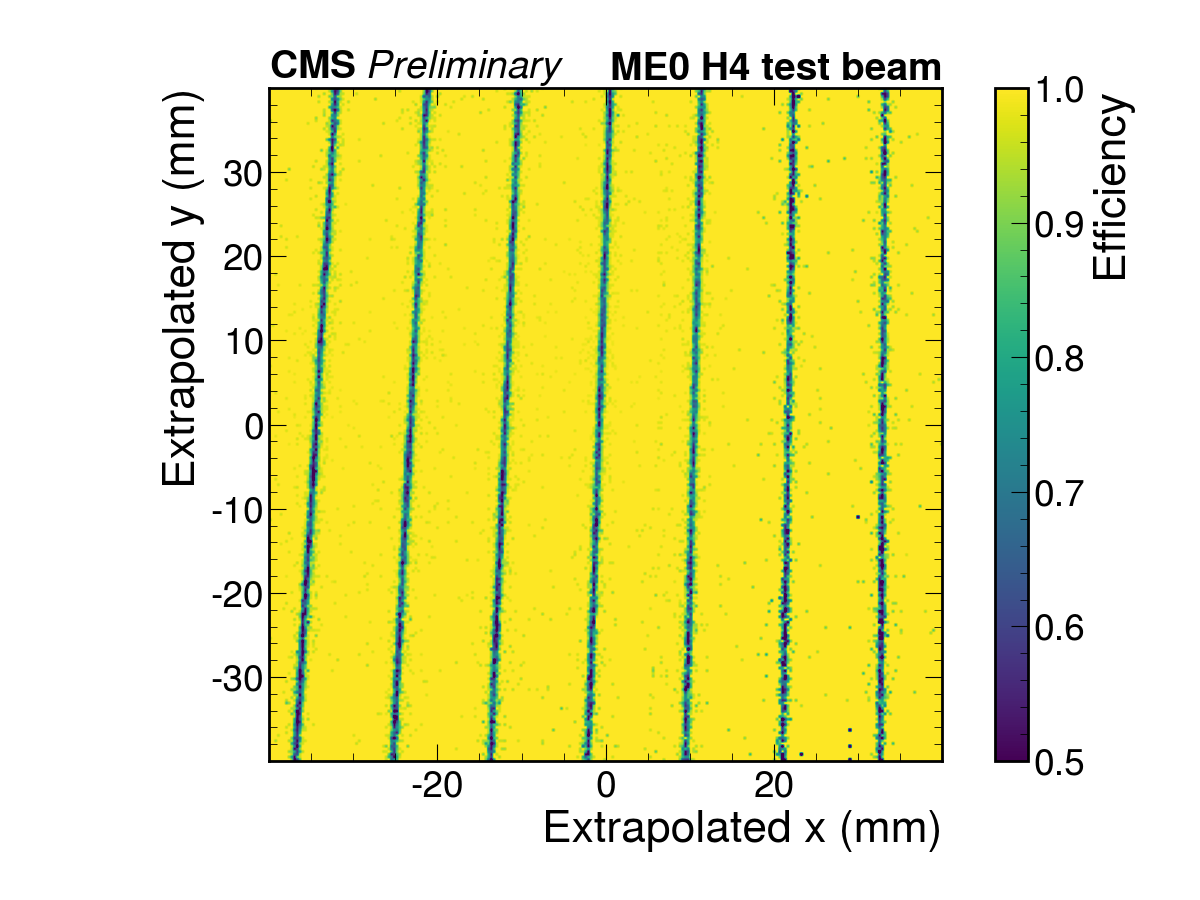}
      \caption{Efficiency maps of the wide side of the ME0 blank-segmented (left) and random-hole segmented (right);
      a clearer distinction of the efficiency losses in the two cases is visible in Fig.\,\ref{fig:me0_profiles}.}
      \label{fig:me0_maps}
  \end{figure}

\begin{figure}[tb]
      \centering
      \includegraphics[width=.495\textwidth]{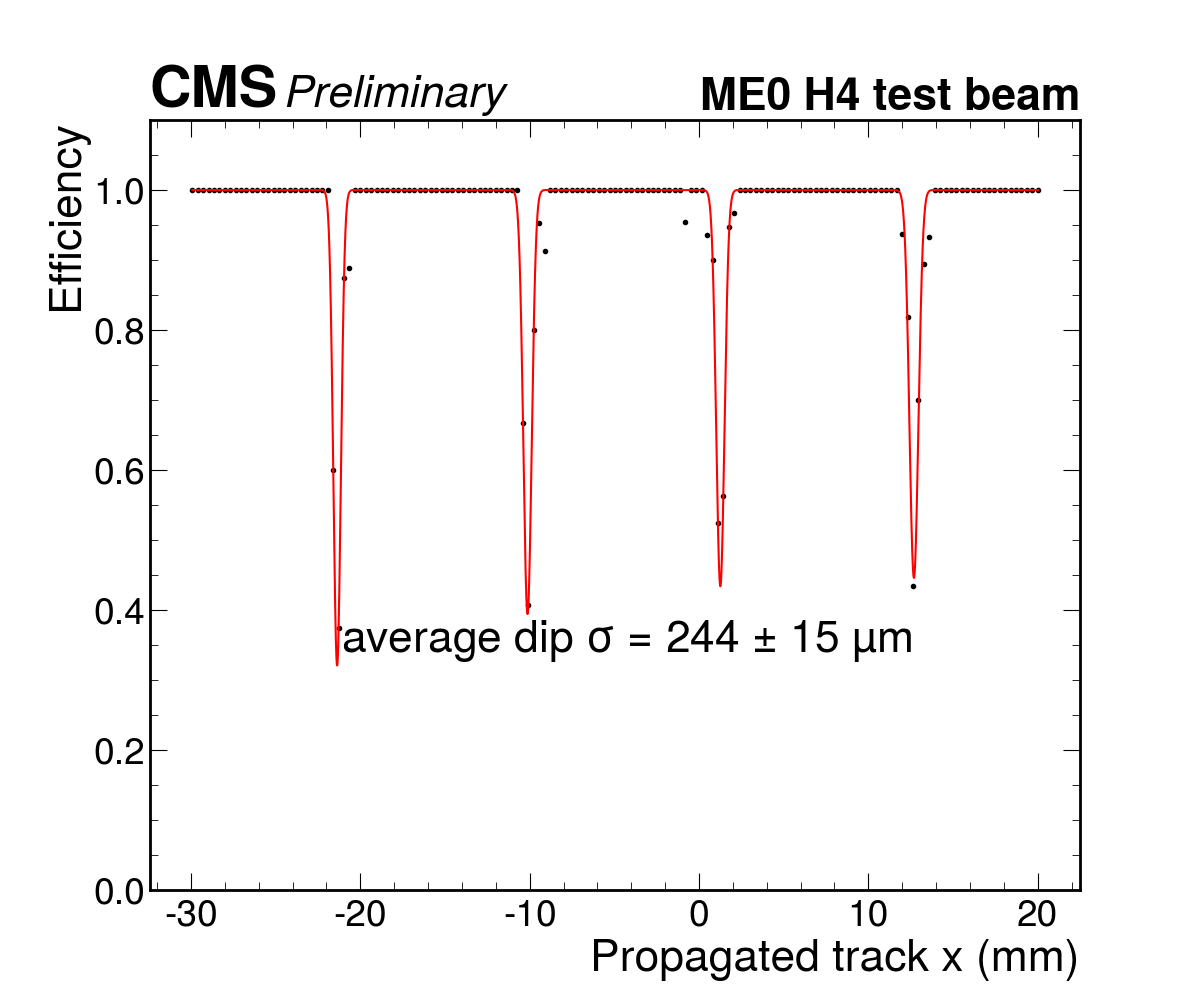}
      \includegraphics[width=.495\textwidth]{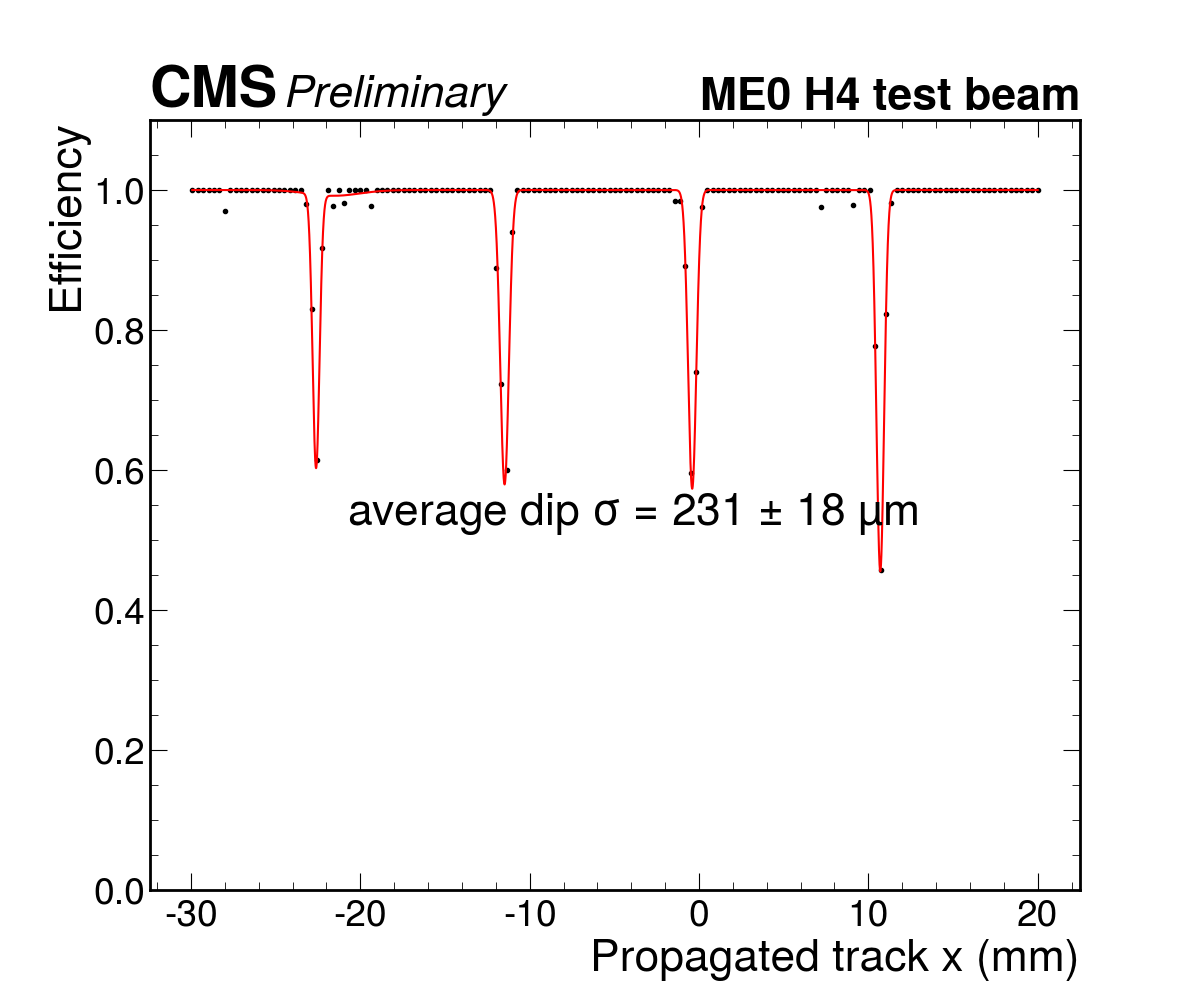}
      \caption{Efficiency profiles of the wide side of the ME0 blank-segmented (left) and random-hole segmented (right).}
      \label{fig:me0_profiles}
  \end{figure}

  Fig.\,\ref{fig:me0_maps} compares the efficiency maps of the blank-segmented ME0 and the random-hole segmented ME0, in the $10\times\SI{10}{\centi\m\squared}$ area covered by the tracker.
  The two maps show the profile of the GEM foil segmentation; 
  the efficiency losses in the dead regions of the random-hole segmented prototype appear lower than in the blank-segmented detector.
  A slice of the efficiency profile cut along the y axis (Fig.\,\ref{fig:me0_profiles}) show the efficiency loss in the blank-segmented detector reaches up to 75\%, while the loss in the random-segmented prototype is limited to about 40\%.

\section{Conclusions}

  The random hole segmentation is a manufacturing technique for sectorized GEM foils that simplifies the etching process with respect to the traditional blank segmentation.
  An additional benefit of the random hole segmentation is the partial recovery of the efficiency loss in the segmentation dead areas, which makes it a potential interest for large-area detectors such as the GEM upgrade of the CMS muon spectrometer.

  A triple-GEM prototype with random hole segmentation has been tested with an x-ray source to measure its gain uniformity, showing a gain loss up to 30\% in correspondence to the dead regions;
  a fine-grain efficiency map obtained with high energy muons in test beam shows that the efficiency loss due to the segmentation is of about 4\% and localized in a width of a few hundreds of µm.

  A comparison in efficiency uniformity between two CMS detector prototypes for the ME0 station -- one with random hole and another with blank segmentation -- showed that in the former the efficiency loss is systematically lower than in the latter.
  With a gain of a few percent in the average efficiency over the entire detector surface, the random hole segmentation is an attractive production choice for the manufacturing of large-area GEM foils and is under consideration for the production of the CMS GEM foils for the ME0 station.





\bibliography{main} 

\end{document}